%% file: main.tex
\documentclass[conference]{IEEEtran}
\IEEEoverridecommandlockouts
\usepackage{cite}
\usepackage{amsmath,amssymb,amsfonts}
\usepackage{algorithmic}
\usepackage{graphicx}
\usepackage{textcomp}
\usepackage{xcolor}
\def\BibTeX{{\rm B\kern-.05em{\sc i\kern-.025em b}\kern-.08em
    T\kern-.1667em\lower.7ex\hbox{E}\kern-.125emX}}
\begin{document}


\long\def\todo#1 { {\bf TODO:} [{\color{blue} #1}] }
\long\def\baf#1{ {\bf RH: } [{\color{gray} \em #1}]}
\long\def\cb#1{ {\bf LJ: } [{\color{gray} \em #1}]}
\long\def\vhpl#1{ {\bf VHL: } [{\color{gray} \em #1}]}
\long\def\veg#1{ {\bf VEG: } [{\color{gray} \em #1}]}


\title{Fair Incentivization of Bandwidth Sharing in Decentralized Storage Networks\\
\thanks{This work is partially funded by the BBChain and Credence projects under grants 274451 and 288126 from the Research Council of Norway.}
}

\author{\IEEEauthorblockN{Vahid Heidaripour Lakhani, Leander Jehl}
\IEEEauthorblockA{
\textit{University of Stavanger}\\
Stavanger, Norway \\
\{heidaripour.l.vahid, leander.jehl\}@uis.no}
\and
\IEEEauthorblockN{Rinke Hendriksen}
\IEEEauthorblockA{
\textit{Ethereum Swarm}\\
Neuchâtel, Switzerland \\
rinke@ethswarm.org}
\and
\IEEEauthorblockN{Vero Estrada-Galiñanes}
\IEEEauthorblockA{
\textit{École Polytechnique Fédérale de Lausanne (EPFL)}\\
Lausanne, Switzerland \\
vero.estrada@epfl.ch}
}

\maketitle
\input{tex/abstract.tex}


\input{tex/introduction.tex}

\input{tex/motivation.tex}
\input{tex/fairness.tex}
\input{tex/related_work.tex}
\input{tex/swarm.tex}
\input{tex/simulation.tex}
\input{tex/discussion.tex}
\input{tex/conclusion.tex}
\input{tex/ack}
\bibliographystyle{IEEEtran}
\bibliography{references.bib}

\end{document}

%% file: tex/abstract.tex
\begin{abstract}
Peer-to-peer (p2p) networks are not independent of their peers, and the network efficiency depends on peers contributing resources. Because shared resources are not free, this contribution must be rewarded. 
Peers across the network may share computation power, storage capacity, and bandwidth. This paper looks at how bandwidth incentive encourages peers to share bandwidth and rewards them for their contribution.
With the advent of blockchain technology, many p2p networks attempt to reward contributions by crypto-assets.
We conduct simulations to better understand current incentive mechanisms, assess the fairness of these mechanisms, and to look for ways to make those incentives more equitable.

\noindent
The following are the primary contributions of this study: (i) We investigate and simulate bandwidth incentives within Swarm, a cutting-edge p2p storage network; (ii) We demonstrate one approach to make the current bandwidth incentives more equitable; (iii) We use the Gini coefficient to define two quantifiable fairness characteristics to evaluate reward sharing in a decentralized p2p storage network.
\end{abstract}

%% file: tex/introduction.tex
\section{Introduction}
A more fair approach to create digital content and reward participants in a community economy is the dream of many developers and cypherpunks, who seek to reshape traditional centralized cloud infrastructures with 
a diverse offer of decentralized systems enabled with blockchain or p2p infrastructure.
These developments invite us to speculate about the changes in the income distribution as we move from a vertical (cloud) to a horizontal (p2p) economic model.   
To illustrate, a dominant global cloud provider has an annual revenue of $\$22$ billions; on the contrary, 
autonomous participants of a p2p economy that provides the same service can expect an annual revenue less than $\$3$ (considering the extremely hypothetical case that every citizen in the world contributes with an equally equipped node to a decentralized network that overpowers the dominant cloud provider, fair incentivization would imply the creation of $7.7$ bn small business (p2p nodes) with $\$2.85$ annual revenue). 
Such a hypothetical grassroot movement would be invaluable for developing and innovating decentralized applications.  
Alongside the real experimentation of promising ideas and novel business models for decentralized applications, a critical systematic scientific review of the novel protocols and system properties can help the open source, blockchain, and p2p communities. 
Woefully, fair incentivization of resource sharing in a p2p network is under-researched. 
Many well-known and heavily researched systems such as Gnutella \cite{gnutellaarticle}, BitTorrent \cite{bittorrent}, and Tor \cite{tor} have surged, especially in the first decade of the $21^{st}$ century, from community efforts.  
More recently, particularly in the area we focus on for this study, developers have started to offer diverse solutions such as IPFS \cite{benet2014ipfs}, Storj \cite{storjwhitepaper}, and Swarm \cite{bookofswarm} as alternatives to the centralized cloud storage owned by a few big corporations. 
These systems still face the same challenges, such as mitigating free-riding and coping with the network churn. However, blockchain and token-based incentives have changed how one can address these problems.
What motivates this study is the need of sustainable token economies to incentivize participants of the Web 3.0 ecosystem.  

Today's web (Web 2.0) has generated massive economic value at the expense of a growing toxic relationship between users, developers, content creators, and the cloud dominant providers. 
Some case examples that illustrate this toxicity are the attention economy, the neglected need for disintermediation, siloed data, and people's resigned acceptance of data surveillance. 
Web 3.0 developers aim at building a decentralized environment with healthier business models and user-centric, self-sovereign solutions that address the aforementioned problems. 
The task is far from trivial, thus, well-established centralized cloud providers are still considered the less risky option to host applications. 
A key question for any serious Web 3.0 developer is: \textit{how can a decentralized infrastructure composed of untrusted network peers offer reliable services with high availability and high performance?} 
Given that the economic value of the growing Web 3.0 ecosystem is not negligible, we believe fair reward mechanisms are key to incentivize reliable peers' resource contributions.  

For example BitTorrent, a network of millions of peers, incentivizes bandwidth contributions with a tit-for-tat mechanisms~\cite{cohen2003incentives}. Such mechanisms ensure that peers receive fair rewards, with respect to their contribution, and prevent free riding.
However, since rewards are only given as access to the service, peers are not incentivized to share resources, when they are not using the system themselves.

More recent networks use more complex and less studied mechanisms based on crypto-tokens to financially incentivize participation in the network. 
IPFS and Swarm are two good examples. Filecoin \cite{filecoin}, an incentive layer in IPFS, encourages peers to contribute and built-in incentives are used by Swarm to reward contributions with the BZZ token.
Crypto-tokens allow to decouple rewards from a peer's usage of the system and can thus incentivize continued contribution.
However, also token based rewards need to disincentivize free riding.

We define two fairness properties which capture these different aspects of disincentivizing free riding and encouraging available contribution.
We show that the degree to which a system fulfills these properties can be measured using the Gini coefficient, a metric to indicate the inequality of income~\cite{gini}.

Since peers are autonomous agents, orchestrating them is the task of a distributed hash table (DHT)~\cite{lua2005survey}. 
In that regard, many decentralized projects chose Kademlia~\cite{10.1007/3-540-45748-8_5} as the coordinator service for its scalability properties. 
We present a tool to analyze reward mechanisms in Kademlia based networks.
As one example, we analyze the bandwidth incentives in Swarm.
Especially, we analyze the effect of parameters in the underlying Kademlia network on the fairness of rewards in Swarm, showing that current parameter settings may not provide optimal motivation to contribute.

The remainder of the paper is structured as follows. Section \ref{sec:motivation} explains why this research is being done, 
defines our fairness properties and presents related work.
In Section \ref{sec:swarm} we give an introduction to Swarm network and its incentives. Section \ref{sec:evaluation} delves into the simulation program and the results. Discussion and future work is presented in Section \ref{sec:discussion}.
The paper concludes with final remarks in Section \ref{sec:conclusion}.
We use the terms \textit{nodes} and \textit{peers} interchangeably throughout this paper to refer to the main entities in a p2p network.

%% file: tex/motivation.tex
\section{Motivation}
\label{sec:motivation}


The adoption of blockchain and crypto-assets in p2p networks provides incentives in exchange for contributions. In that sense, rational peers may be interested in sharing their resources in order to make a profit. 
This payment benefits resource owners while also assisting in the development of more decentralized services. Ecosystems are built to achieve this goal of resource sharing and reward. These ecosystems 
should be profitable enough to entice reasonable peers to contribute (in exchange for rewards), improve network stability, and decrease churn (by staying active in the network). In terms of the storage 
network, an end-user expects to be able to access the stored file in a reliable manner. 
Intuitively, the requirement is to keep service providers sufficiently motivated through incentivization. Understanding how incentives influence 
peer’s network behavior and network properties to design mechanisms that enable a fair and diverse ecosystem is what motivates this study.






%% file: tex/fairness.tex
\subsection{Fairness properties}
\label{subsec:fairness}

For p2p networks that use token rewards to incentivize participation, we define the following two fairness properties:
\begin{enumerate}
    \item[F1] Rewards should be fair (proportional) with regard to a peer's resource contribution to the network.
    \item[F2] Peers willing to provide the same resources should be able to receive an equal share of the reward.
\end{enumerate}
\noindent
F1 states that rewards should be fair, with respect to the resources actually utilized by the system, while F2 states that rewards should be fair, with regard to the resources peers are willing to share. 
As an example, when peers are rewarded for shared bandwidth, F1 implies that rewards should be proportional to the actual number of packets transmitted by a peer; whereas F2 implies that  nodes that make the same bandwidth available to the system, should be able to receive the same reward.

The property F1 is suitable to prevent free-riding, since peers sharing fewer resources will receive fewer rewards. 
F2 on the other hand ensures that peers get an equal opportunity to receive rewards.
We believe that F2 is important to motivate peers to provide continuous service.

We regard properties F1 and F2 not as binary. 
While perfect fairness may be desirable, systems may for example trade some degree of F1 for simpler bookkeeping and reward mechanisms, thus requiring fewer resources to reach a certain level of utility.
Similarly, systems may fail to reach F2 in the presence of low or skewed workload.
We propose to measure F1 and F2 using the Gini coefficient. For a set of values $v_1, ..., v_n$ the Gini coefficient can be computed as 
\begin{equation}
    \frac{\sum_{i=0}^n\sum_{j=0}^n |v_i-v_j|}{2\cdot\sum_{i=0}v_i}
\end{equation}
For F2 we compute the Gini coefficient for the income different nodes get in our simulation. For F1 we first compute the amount of resources used by each peer. We divide this amount by the received reward to get the values $v_i$. We then compute the Gini coefficient, omitting the peers that did not receive any reward.

The Gini coefficient is given in a scale from $0$ to $1$, which in our use case has the following meaning:
\begin{enumerate}
    \item For F1 a coefficient of $1$ means that only one node is rewarded for its contribution. A coefficient of $0$ means that all nodes get the same reward per provided bandwidth.
    \item Similarly, for F2 a coefficient of $1$ means that only one node receives rewards, while a coefficient of $0$ means that all all nodes receive the exact same reward.
\end{enumerate}




%% file: tex/related_work.tex
\subsection{Related work}
\label{subsec:related_work}
Before the emergence of crypto-assets, there was a lot of work on incentives for p2p networks.
Some of the works are aimed at deterring free-riding \cite{cohen2003incentives, jun2005incentives, bharambe2006analyzing, feldman2005overcoming} (to list a few). 
Interestingly Rahman et al. \cite{5502544} already proposed to reward based on the willingness to share resources rather than based on the amount of actual resources shared,
thus focusing on our fairness property F2 rather than F1. 
Reference \cite{daniel2021ipfs} gives an overview of recent decentralized storage networks including their incentives. For various consensus types and perspectives, fair rewarding systems for blockchain-based applications have been investigated \cite{10.1007/978-3-030-32101-7_3, 10.1145/3087801.3087809}, but the problem is that they usually work based on proof of work or proof of stake and are not easily applicable to storage networks. Reference \cite{ghosh2014torpath} proposes TorCoin, an altcoin to reward bandwidth contribution in Tor network.

%% file: tex/swarm.tex
\section{Swarm storage network}
\label{sec:swarm}
Swarm is a p2p network of nodes that collectively provide a decentralized storage and communication service. Its aim is to serve as the storage backbone and virtual server for a wide array of decentralized applications and 
content while maintaining censorship resistance and privacy for the users of the network. The network aims to be self-sustaining and resilient against various attacks through the usage of a crypto-token (BZZ). 
BZZ can be purchased by end-users and exchanged for a service in the network, being storage of data or utilization of bandwidth while uploading or downloading content \cite{bookofswarm}. 

\subsection{Routing in Swarm}
Swarm nodes aim to maintain semi-permanent connectivity with other nodes in the network based on an overlay addressing scheme. The connectivity pattern guarantees that, 
when nodes exchange messages, each edge in the network can be reached with a certain maximum number of hops (logarithmic in the network size). Specifically, the connectivity is based on the topological distance; 
by comparing two Swarm addresses and counting the number of prefix bits which the addresses have in common, we can define their proximity. The furthest away nodes are those nodes with a different first bit, 
the more similar prefix-bits the closer the topological distance between nodes. Based on this distance, a node groups all other nodes in buckets and connects to a number of nodes in each bucket. The zero-bucket 
comprises approximately half of the network and includes all nodes with a different first bit, the first bucket includes about a quarter of the network and includes all nodes with the same first bit and different 
second bit, up to the last bucket (called the neighborhood) which is defined by the proximity at which the node cannot connect to at least four other nodes. 

Figure \ref{fig:routing_bucket} demonstrates an example of a routing table for a node with id $91$. E.g., in the provided routing table in Figure~\ref{fig:routing_bucket}, if a chunk is stored by node with id $245$, then our node will contact one of the four nodes in bucket zero. 
If another chunk is stored by node $64$, then bucket $3$ is contacted.

\begin{figure}[t]
    \centering
    \includegraphics[width=0.95\columnwidth]{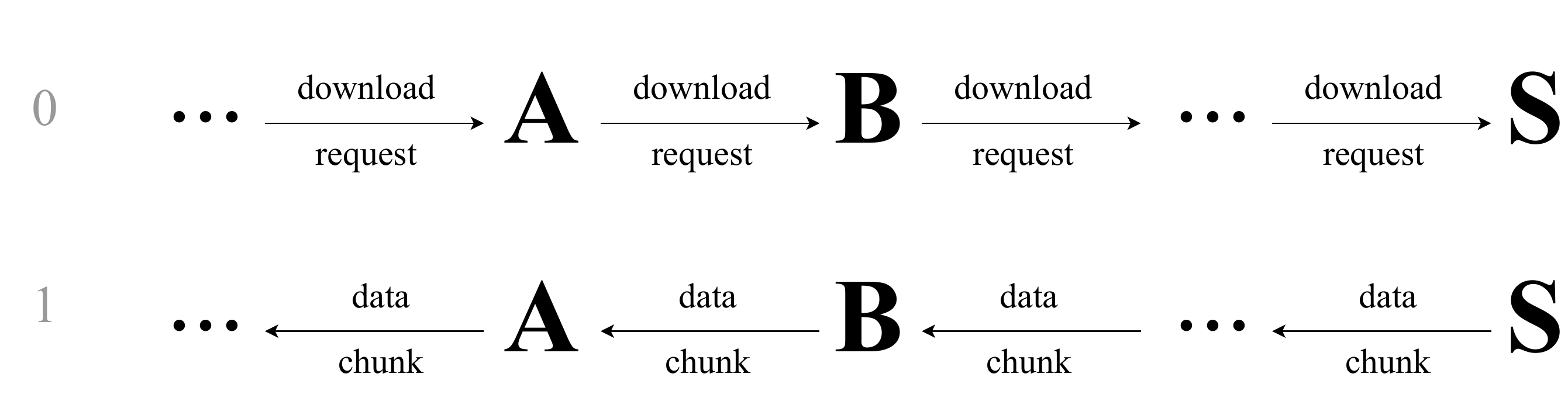}
    \caption{Forwarding Kademlia, which is used by the Swarm network, does not reveal the requestor's identity. In step 0, node A receives a download request and forwards it to the closest possible node while A does not know if the previous node is the originator or just a forwarder like herself. In step 1, node S sends back the data chunk via the same route to the originator but no one knows which node is the final destination.}
    \label{fig:forwarding_kademlia}
\end{figure}

All content in Swarm, fixed size chunks of $4KB$ are addressed on the same address space as nodes. This enables to get a measure of proximity between content and nodes, similar to the proximity 
between two nodes. Content is stored by the closest nodes according to this distance. 
To download a chunk, nodes recursively contact the node closest to the chunks address. The chunk is then forwarded across the same path. This is also shown in Figure~\ref{fig:forwarding_kademlia}.
Upload is done in a similar fashion, where nodes forward the chunk and eventually return a confirmation.

This routing scheme called forwarding Kademlia is different from the Kademlia scheme\cite{10.1007/3-540-45748-8_5}. For the lookup procedure in Kademlia, the node that generated the request repeatedly contacts other nodes for either the chunk, or addresses closer to the chunk. In this way, all involved nodes learn the requester's identity.
Forwarding Kademlia improves privacy and prevents censorship, since nodes cannot distinguish the originator of a request.

\subsection{Incentives in Swarm} 
Incentives are needed in this network for several reasons:
\begin{itemize}
    \item To motivate nodes to join the network through the expectation of a monetary profit.
    \item To ensure that the behavior of each node is in line with the best interest of the network.
    \item To make deliberate attacks on the network costly to mount and persist.
\end{itemize}

Swarm defines two classes of incentives: storage incentives and bandwidth incentives. Storage incentives ensure the allocation of Swarm's finite storage capacity while the bandwidth incentives ensure collaboration 
of nodes in the passing of upload and download messages. Because sharing storage without bandwidth results in files being available online but not retrievable, this study focuses on bandwidth incentives. Sharing bandwidth without storing data allows for the retrieval of popular files at least, which is possible in a network like BitTorrent. Since this paper's focus is on bandwidth incentives, these are detailed further below.

\begin{figure}[t]
  \centering
  \includegraphics[width=0.9\columnwidth]{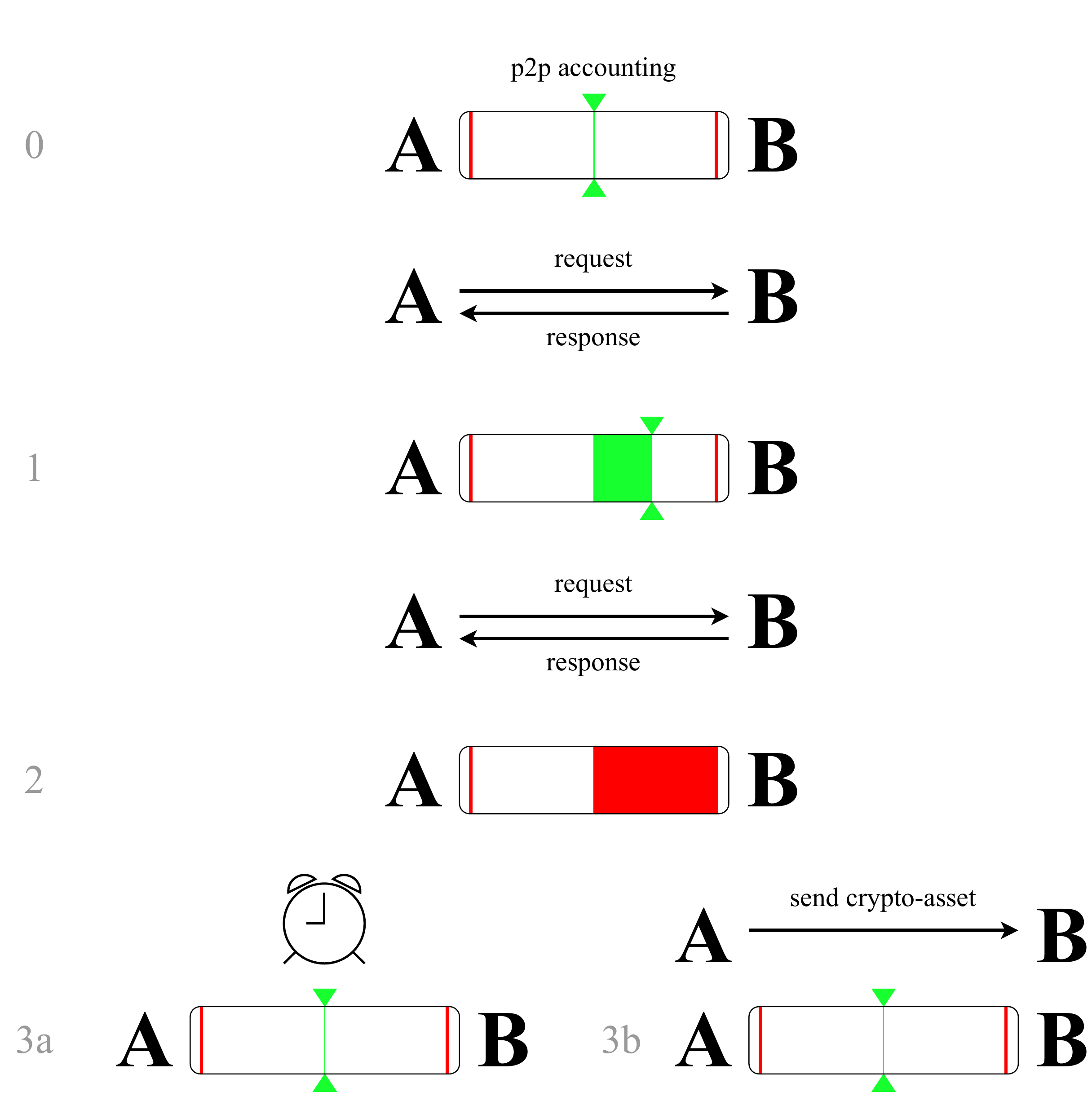}
  \caption{p2p accounting protocol. This diagram shows how peers maintain track of their bandwidth contributions. Starting with a zero balance on both sides at step 0, a period of message exchange follows, leading through step 1, to step n, until the debt on one side hits a threshold. The obligation can be paid off over time, or the creditor can be compensated.}
  \label{fig:swap-swarm}
\end{figure}

At the heart of Swarm's bandwidth incentive system is a system called SWAP: the Swarm Accounting Protocol \cite{tronswap}. In this system, nodes keep track of the relative bandwidth service provided for and consumed by connected peer, in form of \textit{accounting units}. 
Each request for either upload and download is priced respective to the distance between the requester and the destination. Within balance limits, this system enables bandwidth service-for-service exchange. 
If the balance reaches a certain limit, nodes stop serving each others request unless dept is settled by means of payment in BZZ.
Additionally, all balances gravitate continuously to zero via a time-based amortization of balances. 
Thus, nodes may give away a limited amount of bandwidth per time-unit and connection for free. 
This feature allows anybody to request content from Swarm for free--albeit at a slow rate, and to thus encourage adoption. 
The bandwidth incentive mechanism is also shown in Figure~\ref{fig:swap-swarm}. 

When a Swarm node downloads a file, it has to contact the 
one node in its routing table for each of the file's chunks that it does not possess. These nodes contacted by the requester are said to lie in \textit{zero-proximity}.  
The zero-proximity node is the first hop node located in the bucket closest to the destination. 
 
Typically, during a file download, nodes in zero-proximity receive significantly more requests.
Also, nodes in the zero-bucket are only contacted when downloading a file, not when forwarding requests. 
Therefore, the default behaviour implemented in the current Swarm node is to use paid settlement for the requests generated by the originator itself to nodes in the zero-proximity, but to wait for time-based amortization for other requests.


%% file: tex/simulation.tex
\section{Evaluation}
\label{sec:evaluation}

We built a simulation tool to examine the incentives within a p2p storage network. 
Our simulation tool aims to make it easier to analyze complex incentive layers within the networks. The simulation also allows for the measurement of the given fairness properties. 
We currently only modelled Swarm's bandwidth incentives using our tool, but aim to model other networks in the future.

\subsection{Simulator architecture}
The tool's core function is to simulate a p2p network with a specified overlay topology. 
We implement routines to generate an overlay based on Kademlia. Our tool allows to use the same overlay for multiple simulations. This allows us to collect data from runs on multiple machines into a single simulation.
The \textit{cadCAD} simulation engine is used to create the simulation phases. For each step, we simulate the download of a single file, by letting one node request multiple chunks. There are $895$ lines of \textit{Python} code in the current simulation repository.

\begin{figure}[b]
    \centering
    \includegraphics[width=0.9\columnwidth]{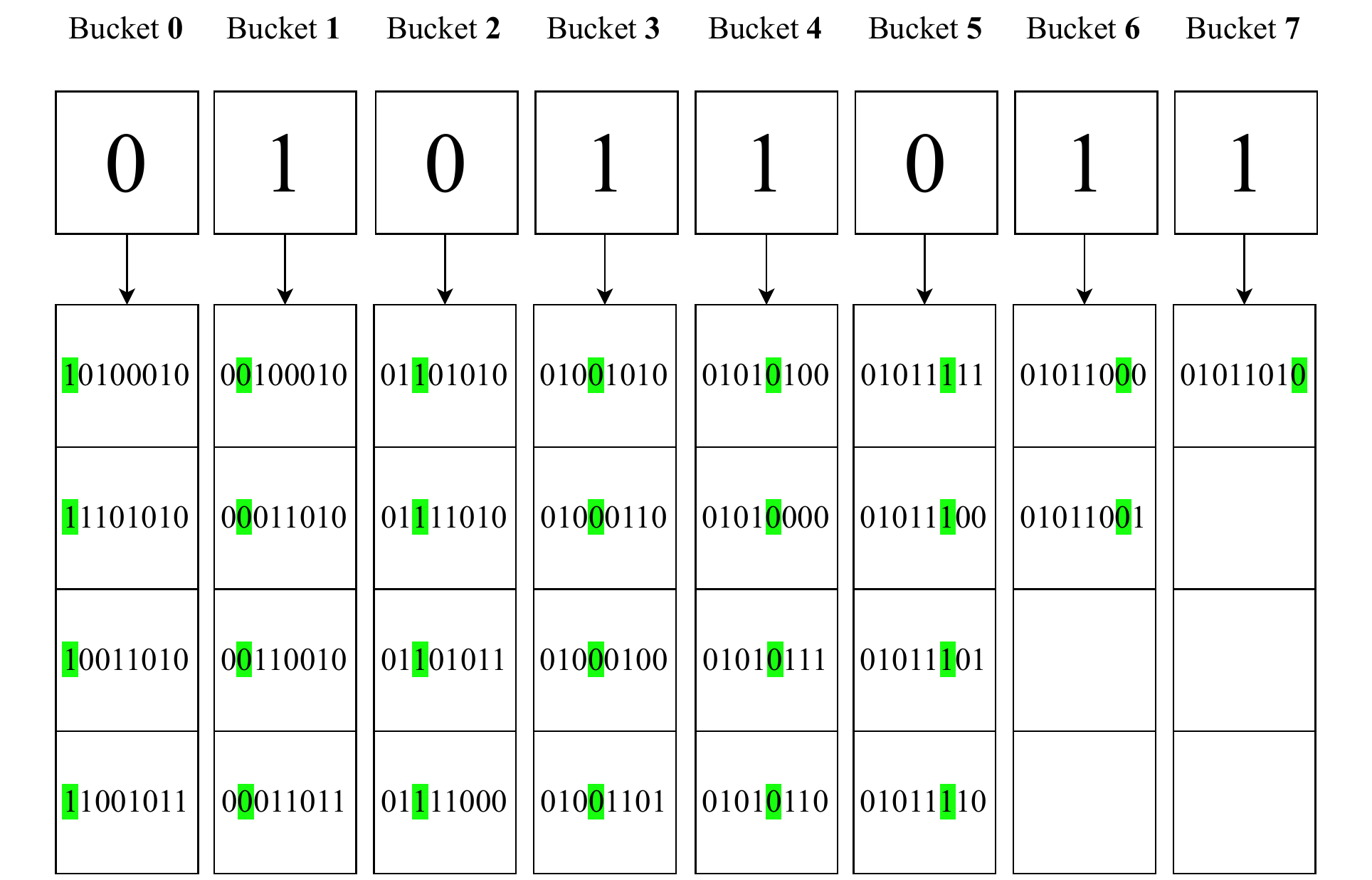}
    \caption{Routing table and buckets for 8 bits of address and k=4.}
    \label{fig:routing_bucket}
\end{figure}

\subsection{Simulation settings}
The simulation environment in this study is designed to model all relevant aspects of the Swarm network. A $1000$-node p2p network is built, by creating a routing table for every node, based on the forwarding Kademlia overlay network. $1000$ nodes generate a reasonably big address space, while simulations are possible on a single machine. This number of nodes was recently used in a Swarm related work~\cite{10.1145/3464298.3493397}. 
The routing tables remain static for the entirety of the experiments, and random numbers are generated using the same seed to ensure consistency throughout all experiments. For the sake of simplicity, we assume that only the node closest to a data chunks address is storing that chunk. The routing tables are divided into $16$ buckets by the address space, which ranges from $0$ to $2^{16}$.
The $16$ address bits are used to form buckets.
The $i$-th bucket of a node contains addresses that have a common prefix of length $i$ with the node's address.
Each bucket contains at most $k$ addresses.
For each given peer, that half of the network's nodes are candidates for bucket $0$, but only $k$ nodes are chosen.
In the Swarm network, the default value for $k$ is $4$, which means that each node can hold the address of four other peers in each bucket. On the contrary, the Kademlia's paper~\cite{10.1007/3-540-45748-8_5} proposed the value 20 for $k$. In our simulation, we compare both settings for k.


To simulate each download request, a random originator generates random chunk requests (all randomness is generated from the uniform distribution.). In Swarm, to download a file, a peer must download all of the files data 
chunks spread throughout the network. To simulate this, 
a single originator requests a random number of chunks,
between $100$ an $1000$.
We call one such step the download of a file.
The addresses of chunks are chosen uniformly at random from the complete address space, $0$ to $2^{16}$.

We performed simulations downloading between $100$ and $10k$ files. 
We perform different simulations where we pick originators uniformly from either $20\%$ or $100\%$ of the nodes, to evaluate the effect of skewed workloads.

To see if there was a difference in the number of accounting units received by a peer during the download process, each simulation is conducted 
with $k = 20$ and $k = 4$. 
We assume, as is the default policy in Swarm, that only nodes within the zero-proximity of a download originator are paid.
The amount of accounting units paid is calculated by using the $XOR$ metric to find the distance to the closest node to the storer.


\begin{figure*}[t]
    \centering
    \includegraphics[width=0.49\textwidth]{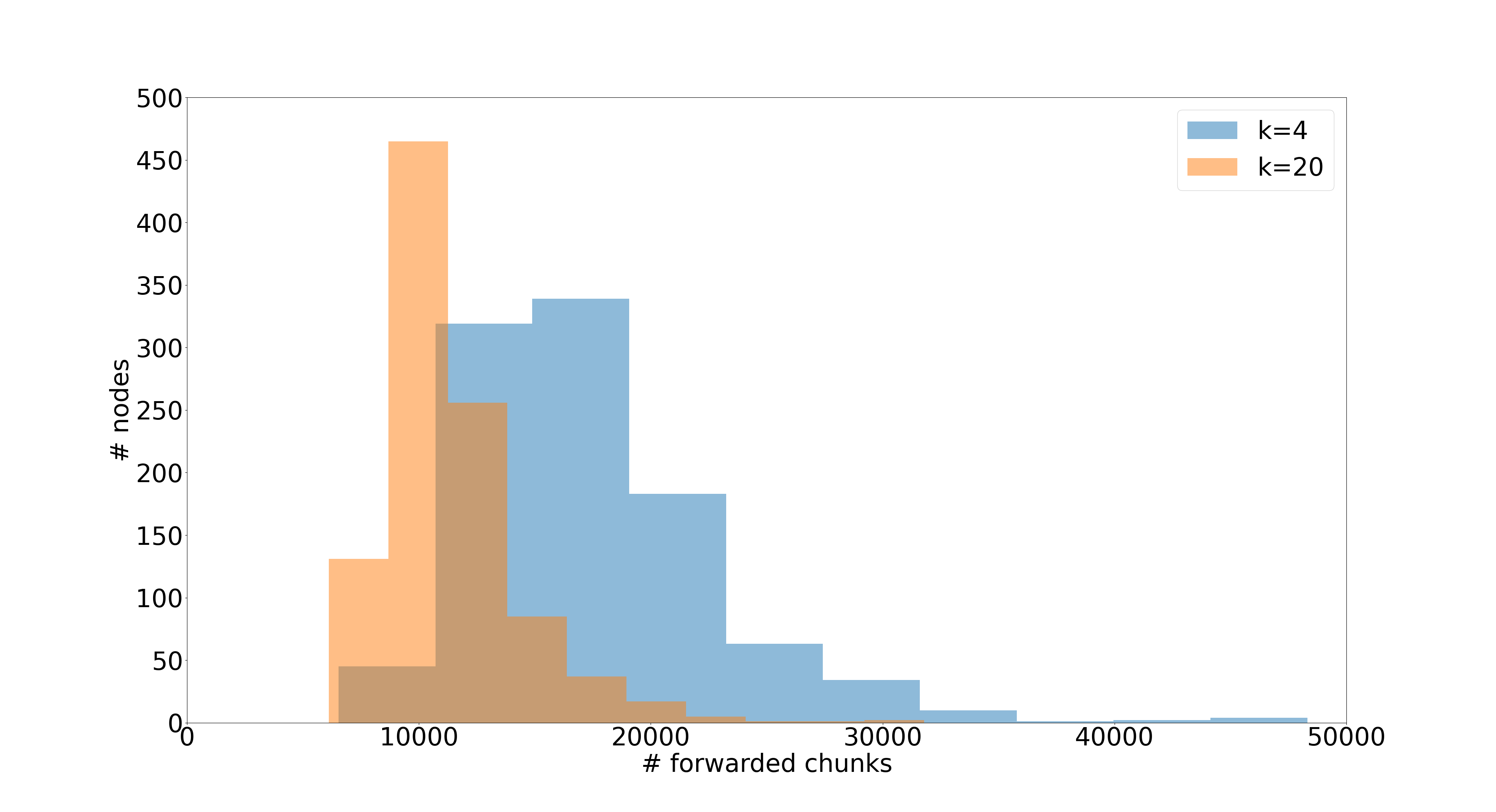}
    \hfill
    \includegraphics[width=0.49\textwidth]{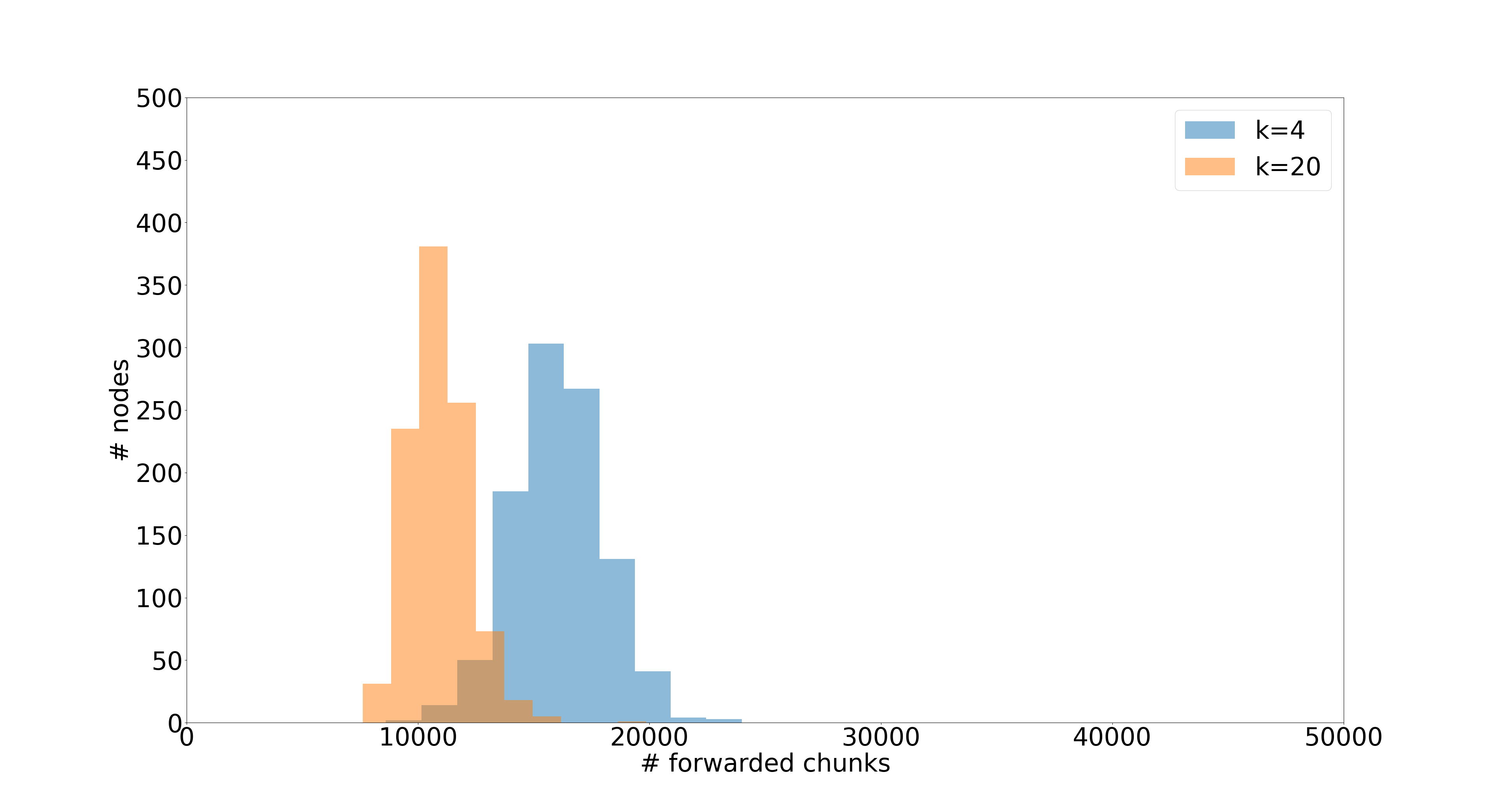}
    \caption{Distribution for the forwarded chunks for 10000 file downloads. Left with 20\% originator, on the right, with 100\% originators.}
    \label{fig:forwarded_chunks}
\end{figure*}

\begin{figure}
    \centering
    \includegraphics[width=0.99\columnwidth]{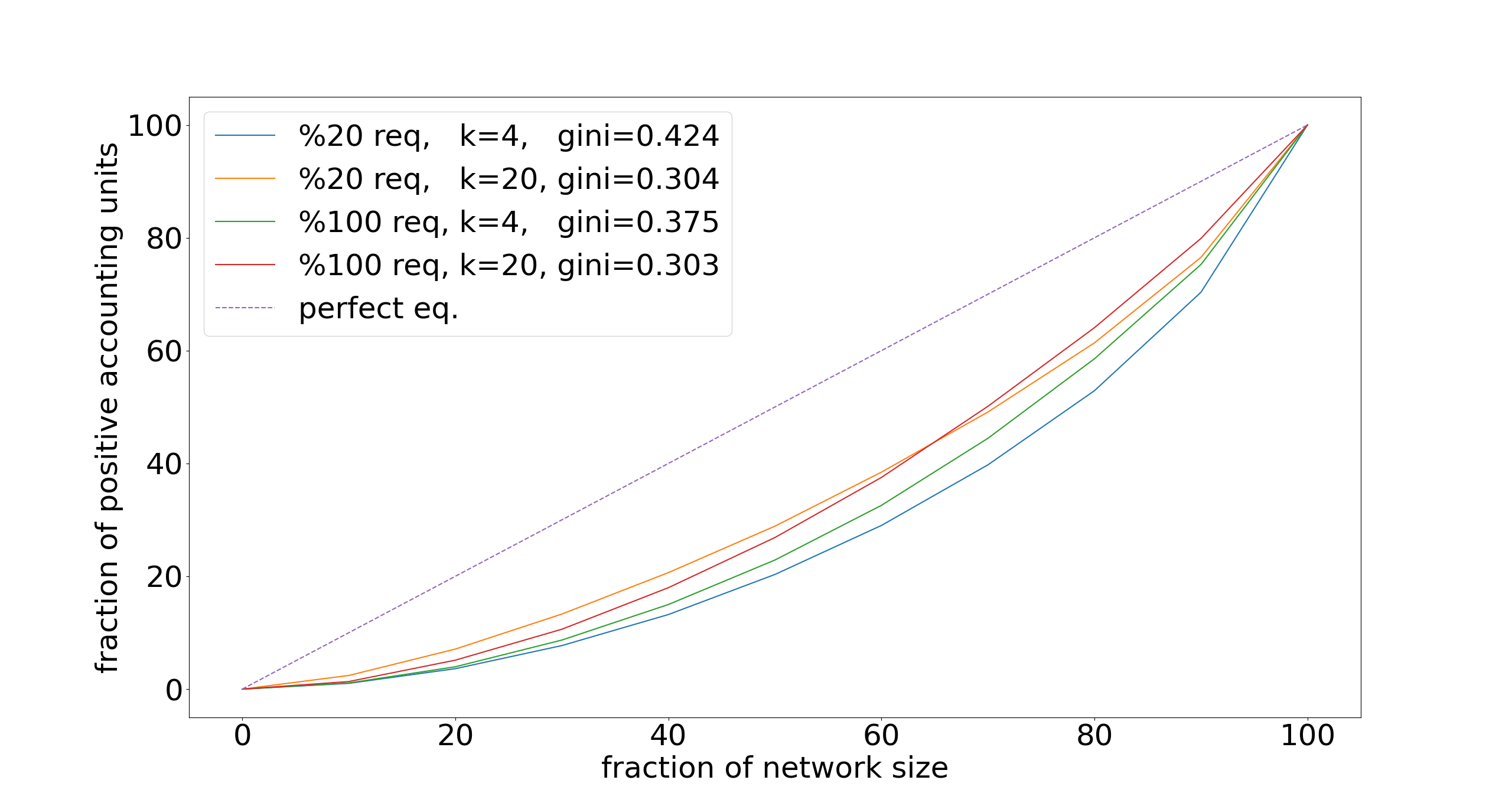}
    \caption{F2 property using Lorenz curve and the Gini coefficient for 10000 file downloads.}
    \label{fig:gini}
\end{figure}

\subsection{Simulation results}
We only include results for runs with $10k$ files downloaded. The other experiments show similar results.

Figure~\ref{fig:gini} shows the Lorenz curves for the distribution of rewarded accounting units and corresponding Gini coefficient, measuring F2 fairness.
Results show that for a bucket size $k$ of $20$, the wealth distribution is more equitable for both 
scenarios ($20\%$ and $100\%$ requests originators). 
For downloading a file, an originator node downloads many chunks. 
For downloading a chunk, only nodes located in downloader's routing table get paid. With larger $k$ the routing table is bigger and payment spreads among more nodes, while having only $4$ options, allows some nodes to accumulate a larger part of the accounting units.
For $k=4$, rewards are also distributed even more unevenly for $20\%$ request originators. 

Table \ref{tab1} shows the average chunks that were transmitted. When $k=20$, the average sent chunks are lower 
for the same download amount, implying that less bandwidth is consumed. Also, when all nodes in the network operate as originators, the average sent data chunks are smaller, which is acceptable because more uniformly distributed originators result in fewer hops to the destination.

Figure \ref{fig:forwarded_chunks} depicts the distribution of forwarded chunks in the $10k$ download experiment. Only $20\%$ of peers behave as originators and produce download requests 
on the left, with $100\%$ of peers acting as originators on the right. 
Frequency on the $y$ axis shows how many chunks are forwarded by individual node, e.g., with $k=20$, more than $400$ out of $1000$ nodes forward approximately $10000$ chunks. 
On the left figure, the area under $k=4$ is $1.6$x bigger than the area for $k=20$, and $1.25$x on the right hand side, indicating that $k=20$ uses less bandwidth on both cases. The results also show that with $20\%$ request originators, bandwidth consumption is distributed more uneven, with many peers using twice the average bandwidth.

\begin{figure}[t]
    \centering
    \includegraphics[width=0.99\columnwidth]{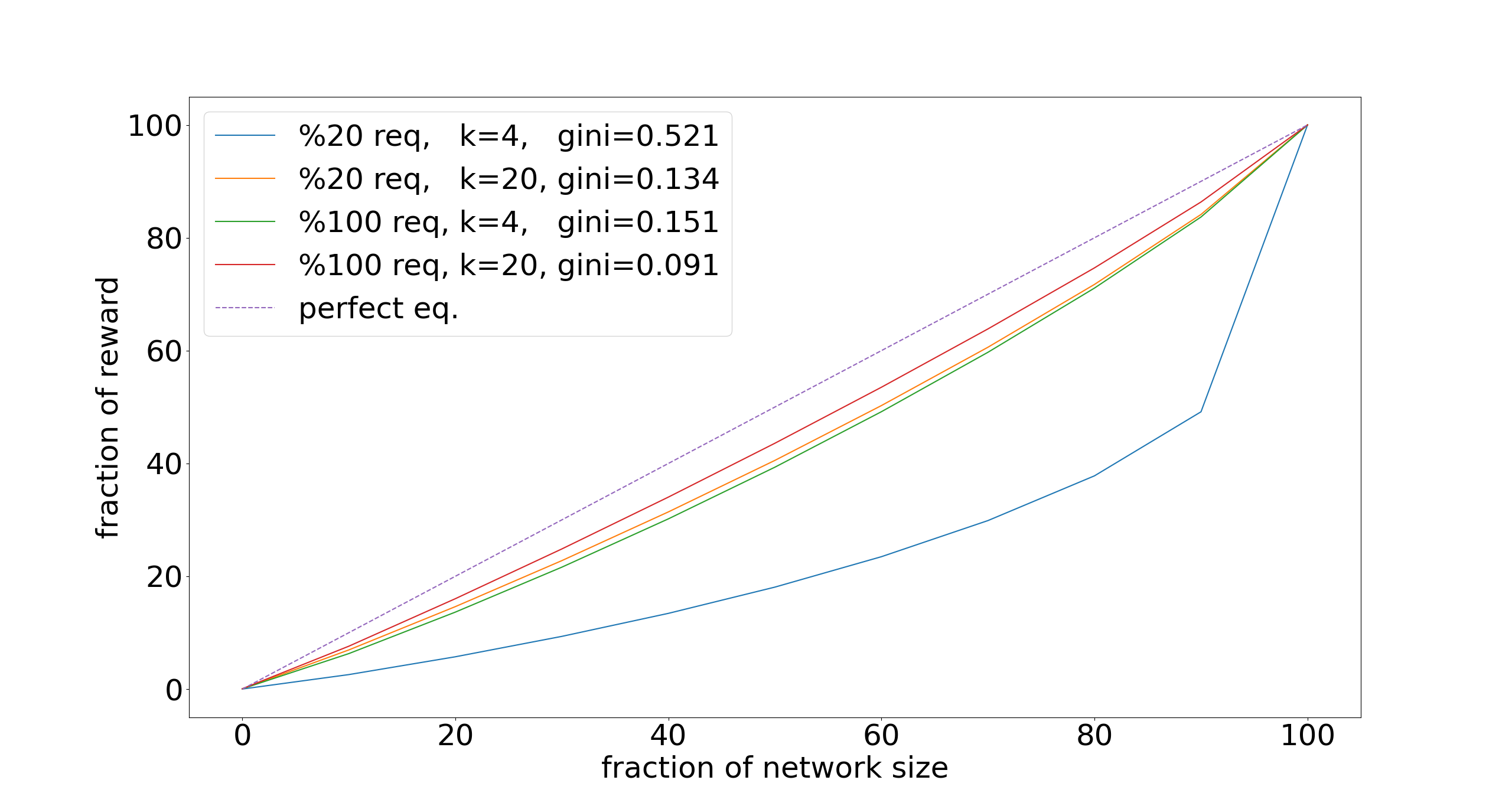}
    \caption{Lorenz curve and Gini coefficient for correlation of total forwarded chunks and forwarded chunks as the first hop.}
    \label{fig:gini_corr}
\end{figure}

Figure \ref{fig:gini_corr} demonstrates the F1 property. For this Figure, we compute the relation of chunks peers serve, to the chunks served as zero-proximity, which peers receive payment for.
The Figure shows Lorenz curves and Gini coefficient for these numbers.
We only consider nodes that actually receive payment. With $k=20$ and $100\%$ of originators, the result reveals a very close number to entire equity. On the other hand, with $k=4$ and $20\%$ request originators, nodes receive very uneven rewards for the provided bandwidth.

\begin{table}
\begin{center}
\caption{Average forwarded chunks for the experiment with 10k downloads}
\begin{tabular}{c|c|c|}
\cline{2-3}
\multicolumn{1}{l|}{}                                                                          & \multicolumn{1}{l|}{20\% originators} & \multicolumn{1}{l|}{100\% originators} \\ \hline
\multicolumn{1}{|c|}{\begin{tabular}[c]{@{}c@{}}Average forwarded chunks\\ k=4\end{tabular}}   & 17253                                 & 16048                                  \\ \hline
\multicolumn{1}{|c|}{\begin{tabular}[c]{@{}c@{}}Average forwarded chunks\\  k=20\end{tabular}} & 11356                                 & 10904                                  \\ \hline
\end{tabular}
\label{tab1}
\end{center}
\end{table}

%% file: tex/discussion.tex
\section{discussion and future work}
\label{sec:discussion}
In a p2p network like Swarm, changing the $k$ value in the overlay network has three basic outcomes:
\begin{itemize}
    \item Increasing $k$ means that each peer has more open connections, which translates to a higher cost for keeping those connections updated.
    \item A peer will have more recipients in the payment channel and issue more payment transactions, resulting each recipient receiving a smaller amount, while the transaction cost for receiving the reward might be more than the reward amount.
    \item With more nodes in each bucket, a peer maintains track of p2p time-based amortization channels with more peers.
\end{itemize}

With the simulation, we demonstrated that with $k=20$, the Gini coefficient approaches a smaller value, but we did not identify the produced overhead in terms of extra bandwidth consumption. There should be a trade-off between the quantity of overhead generated and the amount of money received. In a first thread of future research, we extend our simulation tool in a way that measures the overhead generated. Also, it is interesting to see what happens in payment distribution if we only increase the $k$ for a particular bucket, e.g., bucket zero. Moreover, adding content popularity and caching policies can also have an impact on time-based amortization due to the reduced number of forwarded requests.

For the duration of the experiment, it is assumed that all peers will adhere to the protocol and that no one will attempt to deviate. For example, in this work we assume that nodes are not free-riders, nodes always pay to the zero-proximity node. In a second thread of future work, we will consider what happens when some peers misbehave. An interesting question arises here: \textit{What happens to F1 and F2 properties?}


While creators of these networks claim that the storage incentive makes up the majority of the profit for peers contributing to the network, having not just the bandwidth incentives simulated but also the storage incentives appears needed to complete the simulation and gain a better understanding of a p2p storage network.
 

%% file: tex/conclusion.tex
\section{Conclusion}
\label{sec:conclusion}
In this paper, using our simulation program, we investigated the bandwidth incentives in the growing decentralized p2p Swarm storage network. We propose two fairness properties and measure them in our simulations.
Simulation results show that using a larger bucket size $k$ for the overlay network results in a more fair rewarding system, especially if the system is operating under a skew workload. 
We observed a $7\%$ decrease in the Gini coefficient for F2 and a $6\%$ reduction in the Gini coefficient for F1.

%% file: tex/ack.tex
\section*{Acknowledgment}
The authors wish to thank the anonymous reviewers for their constructive feedback.